\documentclass[twocolumn,english,aps,prl,superscriptaddress,showpacs]{revtex4}
\usepackage{graphicx}
\usepackage{babel}

\begin{document}

\title{Quantum Hall fractions in rotating Bose-Einstein condensates}

\author{N.~Regnault}
\email{Nicolas.Regnault@lpmc.ens.fr}

\author{Th.~Jolicoeur}
\email{Thierry.Jolicoeur@lpmc.ens.fr}

\affiliation{LPMC-ENS, D\'epartement de Physique, 24, rue Lhomond,
75005 Paris, France}


\begin{abstract}
We study the Quantum Hall phases that appear in the dilute limit
of rotating Bose-Einstein condensates. By exact diagonalization in
a spherical geometry we obtain the ground-state and low-lying
excited states of a small number of bosons as a function of the
filling fraction $\nu$, ratio of the number of bosons to the
number of vortices. We show the occurrence of the Jain principal
sequence of incompressible liquids for $\nu = 1/2, 2/3, 3/4, 4/3,
5/4$ as well as the Pfaffian state for $\nu =1$. The collective
excitations are well described by a composite-fermion scheme.
\end{abstract}

\pacs{03.75Kk, 05.30.Jp, 73.43.Cd, 73.43.Lp}

\maketitle



Bose-Einstein condensates in dilute atomic gases offer a unique
opportunity to investigate the physics of vortex matter when they
undergo rotation\cite{Matthews99,Haljan01}. Indeed, recent
experiments \cite{Madison00,Abo01} have observed the appearance of
large vortex arrays at sufficient high angular velocity $\omega$.
In addition to this phase akin to the Abrikosov lattice of type-II
superconductors, there is the possibility that at larger $\omega$
the lattice melts \cite{Sinova02} and is replaced by a quantum
Hall liquid. Consider a trap with strong confinement in the $z$
direction such that the system is effectively two-dimensional
(2D). Then if the rotation frequency is tuned to the
characteristic frequency of the harmonic confining potential in
the $xy$ plane, the bosons feel only the Coriolis force and the
system is equivalent to 2D charged bosons in a magnetic field,
i.e. the conditions of the quantum Hall effect. In this regime, it
has been pointed out \cite{Cooper99,Wilkin00} that the celebrated
Laughlin wavefunction is the exact ground state for the filling
fraction $\nu =1/2$ where $\nu$ is the ratio of the number of
bosons to the number of vortices. Some of the excitations above
this ground state are quasiparticles with fractional statistics
which may eventually be probed by laser manipulations
\cite{Paredes01}. Investigations by exact diagonalization have
given evidence \cite{Cooper01} for even more exotic states of
matter\cite{Moore91,Greiter92}, some involving parafermionic
wavefunctions introduced in the context of the fractional quantum
Hall effect for fermions \cite{ReadRezayi}.


In this Letter we investigate the quantum Hall states of bosons as
a function of the filling $\nu$ by use of exact diagonalizations
in the spherical geometry \cite{Haldane83,Haldane85}. This allows
to separate bulk from edge excitations. We show the appearance of
the Bose analog of the Jain principal sequence of fractions, $\nu
=\frac{n}{n+1}, \frac{p}{p-1}$. The excited states show collective
modes well described by a composite fermion picture in which there
is binding of one vortex per boson. We obtain evidence for the
Pfaffian state \cite{Moore91,Greiter92} at $\nu =1$ by displaying
its peculiar half-vortex excitations. For higher fillings, $\nu
\geq 3/2$, we observe some states with properties of the
Read-Rezayi (RR) parafermionic states. However they show no clear
tendency to convergence to the thermodynamic limit.


In the rotating frame \cite{Rosenbusch02}, the Hamiltonian
describing N bosons of mass $m$ is given by~:
\begin{eqnarray}\label{Ham1}
 \nonumber
  \mathcal{H} &=& \sum^{N}_{i=1}\frac{1}{2m}(\mathbf{p}_{i}-m\omega
    \mathbf{\hat{z}}\times \mathbf{r}_{i})^{2}
    +\frac{1}{2}m(\omega_{0}^{2}-\omega^{2})(x_{i}^{2}+y_{i}^{2}) \\
   &+&\frac{1}{2}m\omega_{z}^{2}z_{i}^{2}+\sum_{i<j}^{N}V(\mathbf{r}_{i}-\mathbf{r}_{j}),
\end{eqnarray}
where the $xy$ trap frequency is $\omega_{0}$, the axial frequency
is $\omega_{z}$ and the angular velocity vector is
$\omega\mathbf{\hat{z}}$. In the ultra-cold atomic gases the
interaction takes place through s-wave scattering only and is thus
given by
$V(\mathbf{r})=(4\pi\hbar^{2}a_s/m)\delta^{(3)}(\mathbf{r})$ where
$a_s$ is the s-wave scattering length. For $\omega$ close to
$\omega_0$, the physics is that of charge-$e$ bosons in a magnetic
field $\mathbf{B}=(2m\omega/e)\mathbf{\hat{z}}$, corresponding to
a magnetic length $\ell=\sqrt{\hbar/(2m\omega)}$. There is then a
2D regime in which the boson wavefunction along the $z$-axis is
the ground state of the harmonic oscillator and the interaction is
now given by
$V^{2D}(\mathbf{r})=g\ell^{2}\delta^{(2)}(\mathbf{r})$ (the vector
$\mathbf{r}$ is 2D), with $g=\sqrt{32\pi}\hbar\omega a_s/\ell_z$,
$\ell_z=\sqrt{\hbar/m\omega_z}$ is the confinement length along
$z$. The energy scale of the quantum Hall problem is thus set by
$g$.


We are thus led to study the quantum Hall effect of bosons
interacting through a delta potential in the lowest Landau level
(LL) \cite{Sinova02-2}. To study the vortex liquids that appear as
a function of the filling factor $\nu$, we use the spherical
geometry \cite{Haldane85,Fano86} in which the bosons move on a
sphere of radius $R$ in the magnetic field of a monopole $B=\hbar
S/eR^{2}$ at the center of the sphere, giving rise to 2S+1
cyclotron orbits in the lowest LL. In the thermodynamic limit, the
filling factor $\nu$ is given by $N/2S$. However for the
incompressible liquids there is in general a finite \textit{shift}
in the relation between the number of particles and the flux, i.e.
one has generally $2S=(1/\nu)N-X$. If we have a guess for the
ground state then one can evaluate the shift and check for its
validity against numerical results. For example the bosonic
Laughlin state for $\nu=1/2$ on the sphere is realized for
$2S=2N-2$ by the wavefunction~:
\begin{equation}\label{LaughlinWavefn}
   \Psi_{1/2}=\prod_{i<j}(u_iv_j-u_jv_i)^2,
\end{equation}
where the spinor coordinates $(u,v)$ are given by~:
\begin{equation}\label{spinor}
(u_i, v_i)=(\cos\theta_i/2 e^{i\phi_i/2}, \sin\theta_i/2
e^{-i\phi_i/2}).
\end{equation}
This is an exact zero-energy eigenstate of the present problem
\cite{Wilkin00}. We have conducted Lanczos diagonalizations for
various N and flux 2S to elucidate the nature of the
incompressible liquid states. States can be labelled by their
total angular momentum $L$, contrary to the planar geometry where
only the z component is conserved.

\textit{Jain sequence.} The signature of incompressible states is
the presence of a L=0 singlet ground state separated by a clear
gap from excited states. A typical spectrum is given in fig.~1a
for $\nu =1/2$. In the excited states we observe a well-defined
collective mode which is gapped for all values of $L$. Clear signs
of incompressible liquids are seen for the fractions $\nu =1/2,
2/3, 3/4, 4/3, 5/4$ : some sizes are displayed in figs.~1b-d.
These fractions are the bosonic analog of the Jain sequence
\cite{Jain89} of fractional quantum Hall states. They are
explained by a composite particle picture in which the composite
fermion (CF) is a boson bound to one vortex. Then the integer
quantum Hall effect with n filled CF LLs leads to a fraction at
$\nu =\frac{n}{n+1}$. This state is realized on the sphere for
$2S=(n+1/n)N-(n+1)$ which is exactly what we observe. The
collective mode is then an exciton-like mode obtained by promoting
one CF from the highest occupied LL to the next LL. On the sphere
the maximal $L$ is then given by $L_{max}=N/n+n-1$ in complete
agreement with our results. We estimate the gap to these neutral
excitations by finite-size scaling \cite{Ambrumenil89} : $\simeq
0.09$g ($\nu =1/2$), 0.05g (2/3), 0.04g (3/4). The series of
values for (N, 2S) is \textit{aliased} \cite{Ambrumenil89} with
the sequence of fractions at $\nu =\frac{p}{p-1}$. Indeed, if (N,
2S) matches the fraction $n/n+1$ then it also matches fraction
$p/p-1$ for p=N/n. Hence the same data set points to the presence
of the fractions $\nu =4/3$ and $5/4$ (we do not have enough
points to provide gap estimates).
\begin{figure}[!htbp]
\begin{center}\includegraphics[  width=3.25in,
  keepaspectratio]{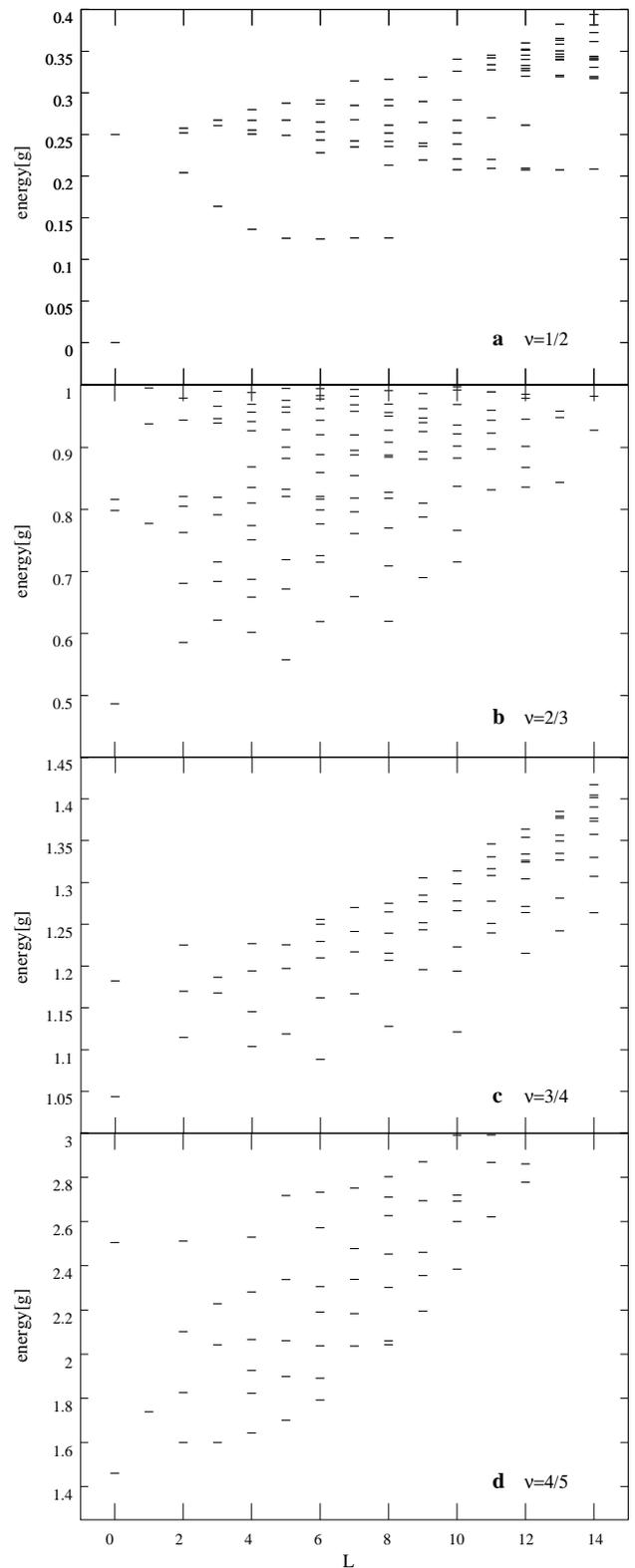}\end{center}
\caption{Energy spectrum for (a) 8 bosons at 2S=14 ($\nu =1/2$),
(b) 8 bosons with 2S=9 (2/3), (c) 12 bosons at 2S=12 (3/4), (d) 8
bosons at 2S=5 (4/5). Energies are in units of $g$ and the
horizontal axis is total angular momentum.}
\end{figure}
The two-particle correlation function $g(r)$ is displayed in
fig.~2 for some states in the CF sequence. For $\nu =1/2$ it is
essentially free of finite-size effects and shows the strong
correlation hole characteristic of a Laughlin ground state. The
state with $\nu =2/3$ no longer vanishes at the origin and hence
has a nonzero ground state energy.
\begin{figure}[!htbp]
\begin{center}\includegraphics[  width=3.25in,
  keepaspectratio]{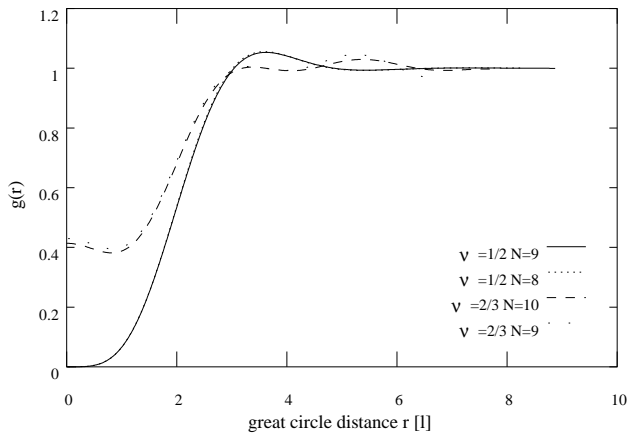}\end{center}
\caption{Two-particle correlation function $g(r)$ as a function of
great circle distance on the Haldane sphere in units of the
magnetic length. The $\nu =1/2$ curve is plotted for sizes N=8,9
and for 2/3 N=9,10.}
\end{figure}
The CF sequence also include the fractions $\nu =3/2$ and $\nu
=2$. For these values we find families of incompressible states in
the (N, 2S) plane but they show no sign of convergence toward the
thermodynamic limit, neither in the ground state energies nor in
the gap values. For these fractions, we have candidates possibly
originating from the Read-Rezayi parafermionic wavefunctions (see
below).

For fillings less than 1/2, the spectrum has many zero-energy
eigenstates that are the quasiholes of the Laughlin state
eq.(\ref{LaughlinWavefn}). This is a special property of the delta
function interaction, the "quasielectron" being gapped. This
obscures the appearance of fractions less than 1/2. If we change
the interaction from pure delta by adding a pseudopotential $V_2$
\cite{Haldane85} in the next allowed partial wave for bosons, i.e.
$\ell =2$, then we find other states from the hierarchy, the
strongest being $\nu =2/5$.
\begin{figure}[!htbp]
\begin{center}\includegraphics[width=3.2in,]{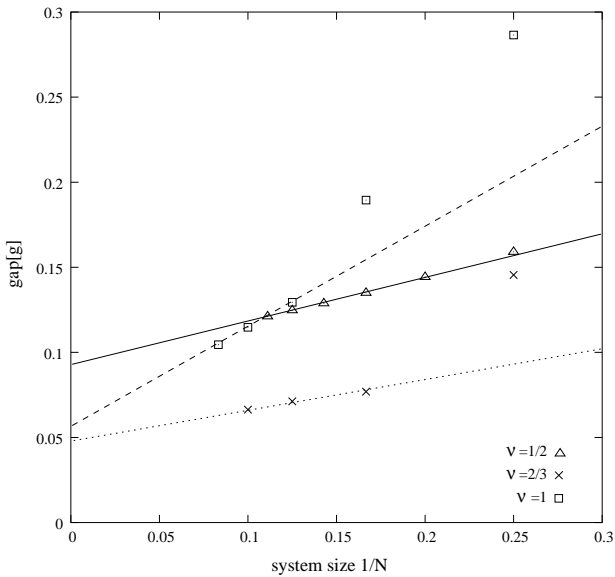}
\end{center} \caption{Gaps for the Bose Jain sequence 1/2 and 2/3 as well as the Pfaffian state.}
\end{figure}

\textit{Pfaffian state.} The filling $\nu =1$ corresponds to the
absence of magnetic field acting upon the composite fermions.
Previous studies \cite{Canright89} are indicative that pairing of
the CFs takes place instead of a Fermi surface. An appealing
wavefunction describing this phenomenon is the so-called Pfaffian
state. On the sphere it can be written as~:
\begin{equation}\label{PfaffianW}
   \Psi_{\nu =1}=\mathrm{Pf}\{\frac{1}{u_iv_j-u_jv_i}\}\prod_{i<j}(u_iv_j-u_jv_i),
\end{equation}
where Pf stands for the Pfaffian which is the antisymmetrized
product of pair wavefunctions \cite{deGennesBook} (a fermionic
version of this state is a good candidate to describe the
enigmatic $\nu =5/2$ quantum Hall state). Calculations of overlaps
between the model Pfaffian wavefunction  Eq.(\ref{PfaffianW})and
the exact ground state suggest
 that it describes the physics of bosons at $\nu =1$ in toroidal
and disk geometry \cite{Cooper99,Wilkin00,Cooper01}. The bosonic
Pfaffian state is realized on the sphere for 2S=N-2 for all N
even. Our calculations lead to incompressible states at these
special values for N=4,6,8,10,12. There is a clear gap that
extrapolates smoothly to $\approx 0.05$g. It appears on fig.~4a
for N=12. This state has charged excitations that are different
from those of a Laughlin fluid. If we add or remove one flux
quantum, then \textit{two} quasiparticles are created, leading to
a set of low-lying states with an alternate even-odd character~:
see fig.~4b. This is consistent with the spectrum for two
identical particles with repulsive interactions. This is observed
for all accessible sizes. We consider this as a proof that the
physics is different from that of the CF sequence and is the
hallmark of the Pfaffian state \cite{Greiter92}.
\begin{figure}
\begin{center}\includegraphics[width=3.25in]{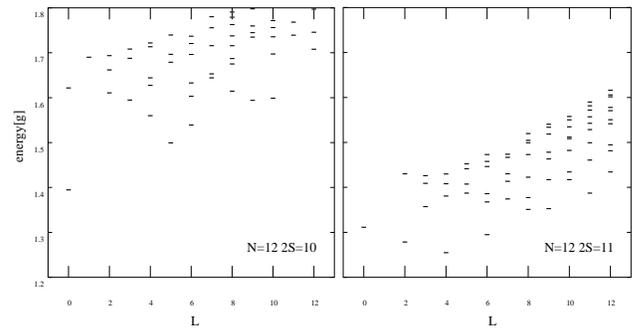}
\end{center} \caption{(a) Spectrum at the Pfaffian matching condition
N=12, 2S=10, with a collective mode above an isolated singlet
ground state (b) with one extra flux quantum, two quasiparticles
give rise to a degenerate set of states L=0,2,4,6.}
\end{figure}
The correlation function is shown in fig.~5. It has now a hump at
the origin possibly due to the pairing of the CFs. The energy of
this state is lowered by a correlation dip (instead of a hole)
which appears at some characteristic radius of order one magnetic
length. For larger separation, $g(r)$ approaches 1 but with a
characteristic length scale which is definitely larger than that
occurring in fig.~2. The correlation length of Pfaffian is larger
than for the Jain-like fractions.
\begin{figure}[!htbp]
\begin{center}\includegraphics[width=3.2in,]{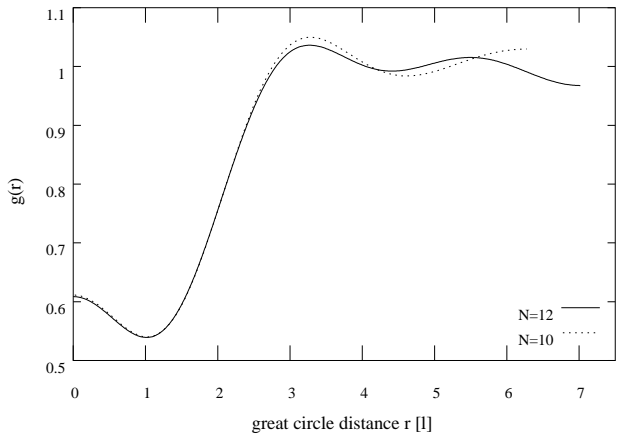}
\end{center} \caption{The two-particle correlation function for the exact ground state
at the Pfaffian point 2S=N-2. It has an extra concentration of
bosons at the origin.}
\end{figure}

\textit{Read-Rezayi states.} For larger fillings it has been
suggested by Cooper et al. \cite{Cooper01} that fractions occur at
$\nu = k/2$ and are well described by the Read-Rezayi
parafermionic wavefunctions. These functions involve clustering of
$k$ particles and are a generalization of the Pfaffian which
corresponds to simple pairing, i.e. $k=2$. On the sphere they are
realized for $2S=\frac{2}{k}N-2$. There are possible candidates at
$\nu =3/2$ for N=6,9,12,15, $\nu =2$ for N=8,12,16 and the
fraction $\nu =5/2$ may be realized for N=15 and 20 (but the gap
is very small). Contrary to the samples belonging to the CF
sequence (they have different fluxes since the shifts are
different between the CF sequence and the RR states) we find no
sign of convergence towards a thermodynamic limit. The gap is
non-monotonous as a function of the size for $\nu =3/2$, and for
$\nu =2$ finite-size effects are very large, preventing any
sensible extrapolation. One possibility is that these states have
very large correlation lengths and are not accommodated on our
largest spheres. This is consistent with the fact that the
correlation function shows very strong oscillations and no hint of
incompressibility.

Evidence for clustering of more particles comes from the
correlation function where we see the same phenomenon as in the
Pfaffian case. The hump at the origin is even more pronounced and
surrounded by a correlation hole. The hump also increases with the
filling albeit we cannot make quantitative statements.

If we increase the number of bosons at fixed flux, then the
spectrum becomes rotor-like : the levels lie on parabolas
described by effective Hamiltonian $\frac{1}{2I}\mathbf{L}^{2}$
and correlation effects disappear.


We have shown the appearance of the Jain principal sequence of
quantum hall fractions in rotating Bose-Einstein condensates. The
composite fermion picture gives a successful account of the
observed fractions as well as their collective mode excitations.
The Pfaffian state is realized at $\nu =1$ as seen from the
special matching of flux and number of bosons as well as its
half-flux quasiparticle. The gaps we estimate from our
diagonalizations are all of the order of $\hbar\omega a_s/\ell_z$.


\begin{acknowledgments}
We thank Yvan Castin and Jean Dalibard for numerous discussions.
\end{acknowledgments}


\end{document}